\documentclass[twocolumn,showpacs,aps,amsmath,amssymb,superscriptaddress]{revtex4-2}  
\usepackage[latin9]{inputenc}
\usepackage{textcomp}
\usepackage{amsmath}
\usepackage{graphicx}
\PassOptionsToPackage{version=3}{mhchem}
\usepackage{mhchem}
\usepackage{xcolor}

\makeatletter

\usepackage{subfigure}\usepackage{epstopdf}\usepackage{dcolumn}
\usepackage{bm}
\usepackage{multirow}
\usepackage{longtable}\usepackage{rotating}\usepackage{threeparttable}
\usepackage{float}

\makeatother

\begin{document}

\title{Rigorous Formalization of Orbital Functionals: Addressing the Noninteracting $v$-Representability Problem}
\author{Neil Qiang Su}
\email[Corresponding author. ]{nqsu@nankai.edu.cn}
\affiliation{Frontiers Science Center for New Organic Matter, State Key Laboratory of Advanced Chemical Power Sources, Key Laboratory of Advanced Energy Materials Chemistry (Ministry of Education), Department of Chemistry, Nankai University, Tianjin 300071, China}

\begin{abstract}
Functionals that explicitly depend on occupied, unoccupied, or fractionally-occupied orbitals are rigorously formalized using Clifford algebras, and a variational principle is established that facilitates orbital (and occupation) optimization as a formal implementation method. Theoretically, these methodologies circumvent the limitations encountered in the original Kohn-Sham and related methods, particularly when the interacting system's electron density does not match that of any noninteracting reference system. This work redefines orbital (and occupation) functionals from a novel perspective, positioning them not merely as extensions of traditional density functionals, but as superior, rigorous alternatives.
\end{abstract}
\maketitle

The Hohenberg-Kohn (HK) theorems fundamentally established the basis for the original density functional theory (DFT) by demonstrating the one-to-one correspondence between the external potential and electron density for nondegenerate ground states, and by establishing a variational principle to compute the ground-state energy directly from densities \cite{HK1964}. However, the HK theorems and their extensions to ensemble degenerate states \cite{Parr1978jcp,Katriel1981ijqc,Levy1982pra,Englisch1983} do not resolve all challenges associated with the implementation and application of functional theories for many-electron systems. Notably, issues related to density \(v\)-representability are crucial---a density is deemed \(v\)-representable if it is a ground-state density for some external potential\cite{HK1964,KS1965}.

Overall, \(v\)-representability issues stem from two critical perspectives: the Hohenberg-Kohn (HK) variational principle\cite{HK1964} and the Kohn-Sham (KS) method\cite{KS1965}. The HK variational principle restricts densities to those of the interacting system under some external potential, thereby introducing the interacting \(v\)-representable (IVR) condition\cite{HK1964}. However, the absence of explicit constraints for IVR densities hampers the variational process, necessitating the inclusion of non-IVR trial densities. Levy addressed this issue through the constrained search formulation \cite{Levy1979pnas}, proposing a universal functional over easily realizable \(N\)-representable (NR) densities---those derivable from any antisymmetric wavefunction. Further efforts on this \(v\)-representability issue have been made \cite{Valone1980jcp,Lieb1983ijqc,Yang2004prl}, with notable contributions from Yang, Ayers, and Wu, who approached the issue through potential functional theory \cite{Yang2004prl}.

Another \(v\)-representability issue emerges in the KS method, which presupposes the existence of a noninteracting reference system governed by an effective potential such that its density matches the ground-state density of the interacting system\cite{KS1965}. This prerequisite, known as the KS \(v\)-representable (KSVR) assumption, is essential for the original KS method and related methodologies \cite{KS1965,Yang2004prl,Talman1976pra,KLI1992pra,Zhao1994pra,van1994pra,Ivanov1999prl,Gorling1999prl,Yang2002prl,Wu2003jcp,Kummel2003prl,Bulat2007jcp}, facilitating the simplification of the variational determination of density and energy. However, the KSVR assumption reveals its limitations when the interacting system's ground-state density is non-KSVR, potentially corresponding to an excited state or failing to match any stationary state of a noninteracting system \cite{Levy1982pra,Yang2004prl,Lee2019prl,Jin2020fd,Yu2021jcp,Su2021pra,DFT2022PCCP}. This KSVR problem thus compromises the reliability of computational results. Moving beyond traditional functionals like the generalized gradient approximations (GGAs) \cite{B88,LYP,PBE,Sun2015prl,LYPr} and hybrid functionals \cite{Becke1993jcp,Adamo1999jcp,Ernzerhof1999jcp,Xu2004,Zhao2008acr,Iikura2001,
Savin1996,Yanai2004,Chai2008jcp,HSE,Garrick2020prx}, emerging functionals that explicitly depend on occupied, unoccupied, or fractionally-occupied orbitals demonstrate superior performance \cite{Zhao2004jpca,Grimme2006jcp,Zhang2009pnas,Scuseria2008,Furche2008jcp,
Hebelmann2011,Ren2011prl,van2013pra,Cornaton2014ijqc,Su2016,Tahir2019prb,Zhang2019jpcl,Muller1984rpa,GU1998prl,Gritsenko2005jcp,Sharma2008prb,Piris2021prl,Ai2022jpcl,Gibney2023prl,Chai2012jcp}. Although it remains feasible to adhere to the KSVR assumption and use the optimized effective potential (OEP) method \cite{Talman1976pra,KLI1992pra,Zhao1994pra,van1994pra,Ivanov1999prl,Gorling1999prl,Yang2002prl,Wu2003jcp,Kummel2003prl,Bulat2007jcp}, to obtain density and orbitals for these new functionals, the KSVR problem persists. Beyond the KS method, new implementations such as orbital optimization \cite{Yaffe1976pra,Lochan2007jcp,Pederson2014jcp,Neese2009jctc,Peverati2013jcp} (or equivalently, generalized OEP (GOEP) \cite{Jin2017jpcl}), alongside combined orbital and occupation optimization \cite{Piris2009jcc,Yao2021jpcl,Yao2024arXiv,Cartier2024arXiv}, have emerged. However, the theoretical foundations of these new functionals and implementations require further demonstration. This work aims to provide a solution to the KSVR problem and establish a rigorous theoretical basis for various orbital functionals and orbital (and occupation) optimization implementations.

For an $N$-electron interacting system subject to an external potential $v$ ($\hat{H}_v=\hat{T} + \hat{V}_{ee}+\sum_i^N v(\textbf{r}_i)$), the Levy constrained search formulation determines the ground-state energy through the following variational principle \cite{Levy1979pnas}:
\begin{equation}
E_v = \min_\rho E_v^{\mathcal{L}}[\rho],
\label{eq:Ev}
\end{equation}
where the trial density $\rho$ can be any NR density, and $E_v^{\mathcal{L}}[\rho]$ is the energy functional given by
\begin{equation}
E_v^{\mathcal{L}}[\rho] = \min_{\Psi \rightarrow \rho} \langle \Psi | \hat{H}  | \Psi \rangle,
\label{eq:Levy}
\end{equation}
in which the minimization searches all normalized antisymmetric wavefunctions that yield $\rho$ to identify the minimizing wavefunction $\Psi_\rho^\text{min}$. Upon successful minimization of Eqs. \ref{eq:Ev} and \ref{eq:Levy}, the resultant energy $E_v$ and the corresponding ground-state density and wavefunction, denoted $\rho_v$ and $\Psi_v$, respectively, are obtained.

Indeed, Eqs. \ref{eq:Ev} and \ref{eq:Levy} correspond respectively to the implementation method and the energy functional required. In practice, an effective strategy is essential to search reasonable NR densities for approaching $\rho_v$. The KS method restricts densities to those corresponding to a noninteracting reference system with an effective potential $w_s$, i.e., the KSVR densities $\rho_{w_s}$ \cite{KS1965,PY1989}. This method not only simplifies the search but also leads to a set of orbitals, known as KS orbitals, which are used to generate $\rho_{w_s}$ and to construct both the kinetic energy functional and the exchange-correlation functional in Eq. \ref{eq:Levy} \cite{DFT2022PCCP}. However, since $\rho_v$ is not necessarily a KSVR density, thus potentially rendering the KS implementation ineffective.

This work abandons the traditional use of a noninteracting reference system to address the limitations associated with KSVR densities. Subsequent discussions transition from a generalized Slater determinant to focusing on orbital functionals. Unlike the ground-state wavefunction of noninteracting systems, which is composed of occupied KS orbitals, the determinant in this study is constructed from an arbitrary set of orthonormal one-electron functions. To explore a more comprehensive scenario, hypercomplex orbitals are employed \cite{Su2021pra}:
\begin{equation}
\label{eq:varphi}
\varphi_{n,i}(\mathbf{r}) = \phi_{i}^{0}(\mathbf{r}) + \sum_{\mu=1}^{n}\phi_{i}^{\mu}(\mathbf{r}) e_\mu,
\end{equation}
with $\{\phi_{i}^{\mu}\}$ as real functions and $\{e_1, \ldots, e_n\}$ as a basis in a Clifford algebra of dimension $n$ \cite{CA2019}, satisfying:
\begin{equation}
e_\mu^2 = -1; \quad e_\mu e_\nu = -e_\nu e_\mu.
\end{equation}
The determinant constructed from $\{\varphi_{n,i}\}$ is denoted $\Phi_n$.  At $n=0$, $\{\varphi_{n,i}\}$ and $\Phi_n$ revert to traditional real orbitals and determinant. With this construction, the density for Eq. \ref{eq:Ev} can be obtained via:
\begin{equation}
\langle \Phi_n | \hat{\rho} | \Phi_n \rangle = 2\sum_i |\varphi_{n,i}(\mathbf{r})|^2 =2 \sum_p \lambda_{n,p} |\chi_{n,p}(\mathbf{r})|^2,
\label{eq:trialrho}
\end{equation}
where the last equality was derived in Ref. \citenum{Su2021pra}, and $\{\chi_{n,p}, \lambda_{n,p}\}$ represent hierarchically correlated orbitals (HCOs) and their occupations \cite{Su2022es}. As indicated by Eq. \ref{eq:trialrho}, the density remains invariant under the unitary rotation of $\{\varphi_{n,i}\}$ and of $\{\chi_{n,p}\}$ with the same occupation.

\begin{figure}[h]
 \centering
\includegraphics[width=1\linewidth]{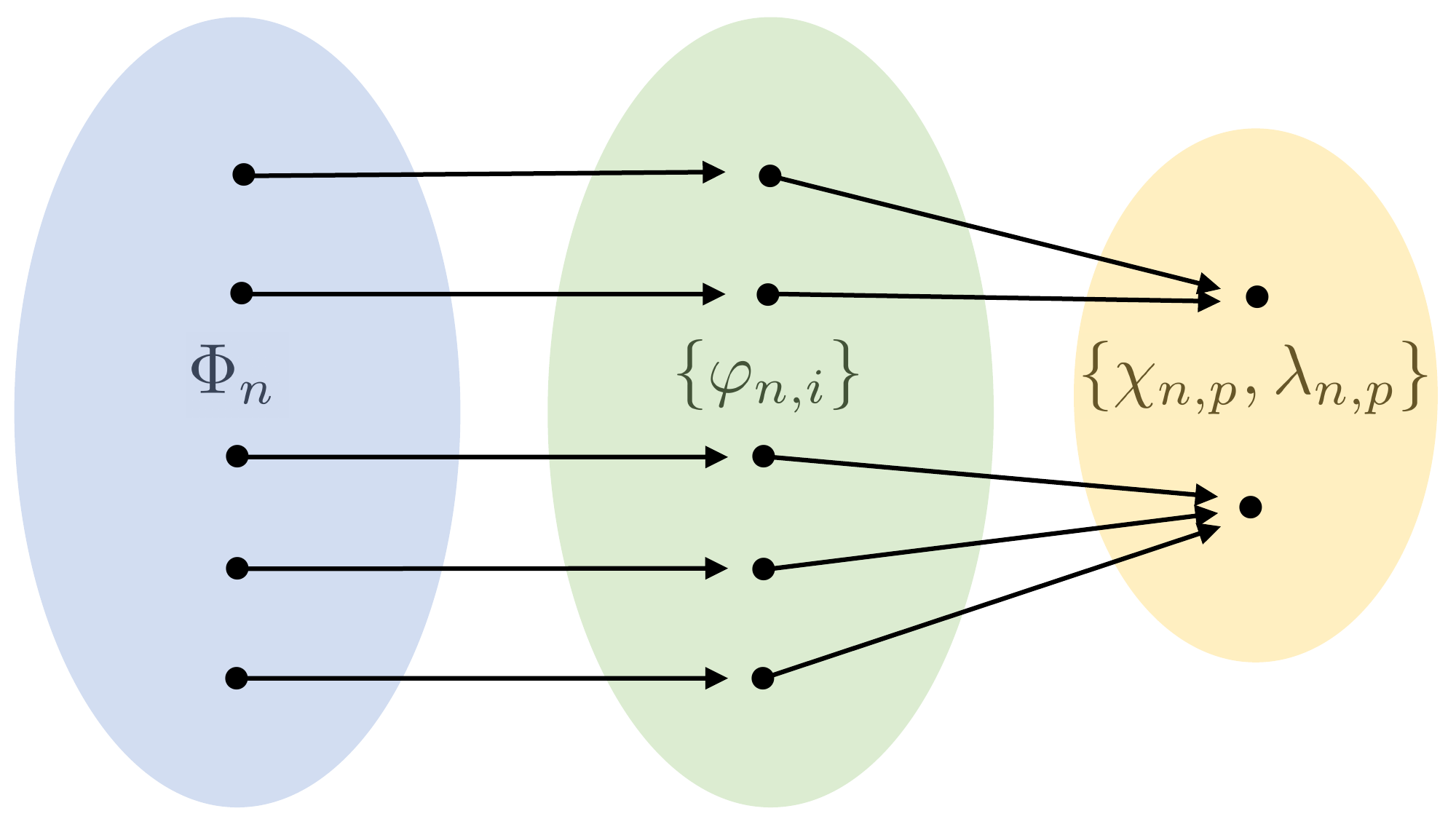}
 \caption{Mappings among the domains of $\Phi_n$, $\{\varphi_{n,i}\}$, and $\{\chi_{n,p}, \lambda_{n,p}\}$.}
 \label{fig:mapping}
\end{figure}

\textbf{Theorem 1:} The mappings among $\Phi_n$, $\{\varphi_{n,i}\}$, and $\{\chi_{n,p}, \lambda_{n,p}\}$ are surjective for a given $n$, as shown in:
\begin{equation}
\label{eq:domain}
\Phi_n \Leftrightarrow\{\varphi_{n,i}\} \Leftrightarrow \{\chi_{n,p}, \lambda_{n,p}\}.
\end{equation}
Specifically, there exists a one-to-one mapping between $\Phi_n$ and $\{\varphi_{n,i}\}$, and a many-to-one mapping between $\{\varphi_{n,i}\}$ and $\{\chi_{n,p}, \lambda_{n,p}\}$, within the unitary rotation of $\{\varphi_{n,i}\}$ and of $\{\chi_{n,p}\}$ with the same occupation. Refer to Fig. \ref{fig:mapping} for visualization.

\textbf{Proof 1:} The proof of this theorem is detailed in the supplementary material \cite{footnote}. \(\square\)

Theorem 1 establishes that the densities derived from $\Phi_n$, $\{\varphi_{n,i}\}$, and $\{\chi_{n,p}, \lambda_{n,p}\}$---collectively referred to as $\rho_n$---share the same domain. The subsequent theorem will delve deeper into the properties of $\rho_n$ and delineate their advantages over $\rho_{w_s}$.

\begin{figure}[htbp]
\centering
\includegraphics[width=\linewidth]{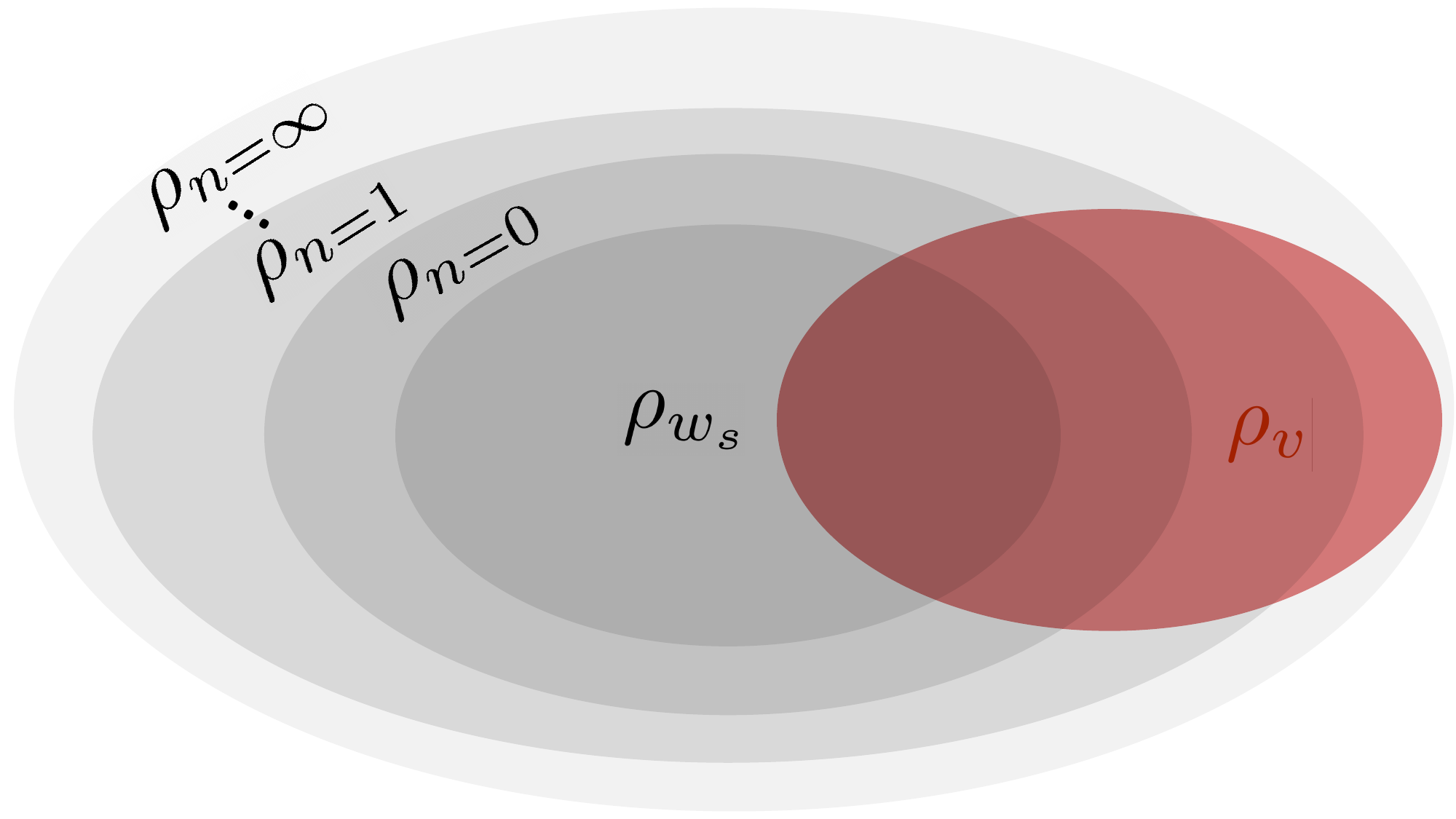}
\caption{The matryoshka-like structure of $\rho_n$ domains across different \(n\) levels, illustrating their relationships with KSVR and IVR densities.}
\label{fig:rho}
\end{figure}

\textbf{Theorem 2:} The domain of $\rho_n$ expands as \(n\) increases, exhibiting a matryoshka-like structure. The relationships among the domains of $\rho_{w_s}$, $\rho_{v}$, and $\rho_{n}$ from \(n=0\) to \(n=\infty\) are shown as:
\begin{equation}
\label{eq:domrho}
\text{dom}\rho_{w_s} \subseteq \text{dom}\rho_{n=0} \subseteq \text{dom}\rho_{n=1} \cdots \subseteq \text{dom}\rho_{n=\infty} \supseteq \text{dom}\rho_{v}
\end{equation}
This indicates that the domain of $\rho_n$ fully encompasses KSVR densities and progressively includes all IVR densities, as depicted in Fig. \ref{fig:rho}.

\textbf{Proof 2:} Initially, consider the relationship between \(\rho_{w_s}\) and \(\rho_{n=0}\). Since the ground state of a noninteracting reference system is typically represented by a Slater determinant, all \(\rho_{w_s}\) inherently reside within the domain of \(\rho_{n=0}\). However, certain \(\rho_{n=0}\) densities are identified as non-KSVR, either corresponding to excited states or not aligning with any stationary state of a non-interacting system, establishing the first relation in Eq. \ref{eq:domrho}. This analysis assumes \(\rho_{w_s}\) is derived from conventional real KS orbitals. Should complex KS orbitals be used, \(\rho_{n=0}\) transitions to \(\rho_{n=1}\), with no additional modifications required.

The matryoshka-like structure of the domains of $\rho_n$ for different \(n\) levels is elucidated by Eq. \ref{eq:varphi}. As \(n\) increases, the degrees of freedom for \(\{\varphi_{n,i}\}\) increase, broadening the domain of \(\rho_{n}\). Moreover, \(\{\varphi_{n,i}\}\) at lower \(n\) can be dimensionally reduced from those at higher \(n\), ensuring that the domain of \(\rho_{n}\) at higher \(n\) fully encompasses that at lower \(n\). Additionally, the constraints on \(\{\chi_{n,p}, \lambda_{n,p}\}\) \cite{Su2021pra,Zhang2023pra} illustrate another aspect: at \(n=0\), \(\{\chi_{n,p}, \lambda_{n,p}\}\) are subject to stringent constraints---\(\{\chi_{n,p}\}\) are orthonormal and \(\{\lambda_{n,p}\}\) are integers, with \(N\) being 1 and the rest 0; as \(n\) increases, these constraints relax, thereby extending \(\rho_{n}\)'s domain.

Regarding the domain of \(\rho_{n=\infty}\), as \(n\) approaches infinity, the constraints on \(\{\chi_{n,p}, \lambda_{n,p}\}\) become minimal, described by \cite{Su2021pra}:
\begin{equation}
\label{eq:cons_hco}
\langle \chi_{n,p} | \chi_{n,q}\rangle = \delta_{pq},
\end{equation}
\begin{equation}
\label{eq:cons_hcoo}
0 \leq \lambda_{n,p} \leq 1, \quad \sum_{p} \lambda_{n,p} = N/2.
\end{equation}
As detailed in Eq. \ref{eq:trialrho}, the density \(\rho_n\) in terms of \(\{\chi_{n,p}, \lambda_{n,p}\}\) is:
\begin{equation}
\rho_n(\mathbf{r}) = 2 \sum_p \lambda_{n,p} |\chi_{n,p}(\mathbf{r})|^2.
\label{eq:trialrho2}
\end{equation}
With these constraints, Eq. \ref{eq:trialrho2} can reproduce any NR density, conclusively establishing that the domain of \(\rho_{n=\infty}\) encompasses that of \(\rho_{v}\). \(\square\)

The above two theorems confirm that $\Phi_n$, $\{\varphi_{n,i}\}$, and $\{\chi_{n,p}, \lambda_{n,p}\}$ are equivalent for generating trial densities. The resulting densities, $\rho_n$, are notable for their scalability, extending their domain with increasing \(n\) to cover a broader range of plausible NR densities. Additionally, as derived from Ref. \citenum{Harriman1981pra}, under the ideal complete basis set limit, $\rho_{n=0}$ can produce any NR density, making the domain of $\rho_n$ uniform across all \(n\). Consequently, $\rho_n$ are termed meta-NR densities, highlighting their ability to transcend traditional density search paradigms. Building on these insights, further discussions will elucidate how $\Phi_n$, $\{\varphi_{n,i}\}$, and $\{\chi_{n,p}, \lambda_{n,p}\}$ serve as fundamental descriptors of the system.

Taking $\{\chi_{n,p}, \lambda_{n,p}\}$ as an example, the following variational principle is constructed:
\begin{equation}
E_{n,v} = \min_{\{\chi_{n,p}, \lambda_{n,p}\}} E_{n,v}^{\text{pw}}[\{\chi_{n,p}, \lambda_{n,p}\}],
\label{eq:Env}
\end{equation}
with the energy functional defined as:
\begin{equation}
E_{n,v}^{\text{pw}}[\{\chi_{n,p}, \lambda_{n,p}\}] = \min_{\Psi \rightarrow \rho_{\{\chi_{n,p}, \lambda_{n,p}\}}} \langle \Psi | \hat{H}_v | \Psi \rangle,
\label{eq:EnvPhi}
\end{equation}
where the superscript pw stands for present work. Eq. \ref{eq:Env} seeks to minimize across the domain of $\{\chi_{n,p}, \lambda_{n,p}\}$, with the density of each trial $\{\chi_{n,p}, \lambda_{n,p}\}$ given by $\rho_{\{\chi_{n,p}, \lambda_{n,p}\}} = 2 \sum_p \lambda_{n,p} |\chi_{n,p}(\mathbf{r})|^2$.

\textbf{Theorem 3:} Assume $\rho_v$ is a meta-NR density for a given $n$. Then, $E_{n,v}^{\text{pw}}[\{\chi_{n,p}, \lambda_{n,p}\}]$ is stationary with respect to variations in $\{\chi_{n,p}, \lambda_{n,p}\}$ when $\rho_{\{\chi_{n,p}, \lambda_{n,p}\}} = \rho_v$, reaching its minimum and providing the ground state energy:
\begin{equation}
E_v = E_{n,v}.
\end{equation}

\textbf{Proof 3:} The variational principle in Eq. \ref{eq:Ev} stipulates that when the trial density equals the ground state density ($\rho = \rho_v$), the Euler-Lagrange equation is satisfied \cite{PY1989}:
\begin{equation}
\frac{\delta E_v^{\mathcal{L}}[\rho]}{\delta \rho(\textbf{r})}\bigg|_{\rho(\textbf{r})=\rho_v(\textbf{r})} = \mu,
\label{eq:EulerLevy}
\end{equation}
where $\mu$ is the chemical potential ensuring the density is NR. For $E_{n,v}^{\text{pw}}[\{\chi_{n,p}, \lambda_{n,p}\}]$, consider small variations within $\{\chi_{n,p}, \lambda_{n,p}\}$'s domain:
\begin{equation}
\frac{\delta E_{n,v}^{\text{pw}}[\{\chi_{n,p}, \lambda_{n,p}\}]}{\delta \{\chi_{n,p}, \lambda_{n,p}\}} = \int \frac{\delta E_{n,v}^{\text{pw}}[\{\chi_{n,p}, \lambda_{n,p}\}]}{\delta \rho_{\{\chi_{n,p}, \lambda_{n,p}\}}(\textbf{r})} \frac{\delta \rho_{\{\chi_{n,p}, \lambda_{n,p}\}}(\textbf{r})}{\delta \{\chi_{n,p}, \lambda_{n,p}\}} d\textbf{r}.
\label{eq:EulerPhi}
\end{equation}
Upon comparing the definitions of $E_v^{\mathcal{L}}[\rho]$ and $E_{n,v}^{\text{pw}}[\{\chi_{n,p}, \lambda_{n,p}\}]$ (namely Eqs. \ref{eq:Levy} and \ref{eq:EnvPhi}), when $\rho_{\{\chi_{n,p}, \lambda_{n,p}\}} = \rho_v$, the above equation simplifies to:
\begin{equation}
\frac{\delta E_{n,v}^{\text{pw}}[\{\chi_{n,p}, \lambda_{n,p}\}]}{\delta \{\chi_{n,p}, \lambda_{n,p}\}} = \mu \int \frac{\delta \rho_{\{\chi_{n,p}, \lambda_{n,p}\}}(\textbf{r})}{\delta \{\chi_{n,p}, \lambda_{n,p}\}} d\textbf{r} = 0,
\label{eq:EulerPhiSolved}
\end{equation}
confirming that $E_{n,v}^{\text{pw}}[\{\chi_{n,p}, \lambda_{n,p}\}]$ is stationary. Furthermore, the following can be obtained:
\begin{align}
E&_{n,v}^{\text{pw}}[\{\chi_{n,p}, \lambda_{n,p}\}]|_{\rho_{\{\chi_{n,p}, \lambda_{n,p}\}} \neq \rho_v} = \min_{\Psi \rightarrow \rho_{\{\chi_{n,p}, \lambda_{n,p}\}}} \langle \Psi | \hat{H}_v | \Psi \rangle  \nonumber \\
& \geq \min_{\Psi \rightarrow \rho_{v}} \langle \Psi | \hat{H}_v | \Psi \rangle = E_{n,v}^{\text{pw}}[\{\chi_{n,p}, \lambda_{n,p}\}]|_{\rho_{\{\chi_{n,p}, \lambda_{n,p}\}} = \rho_v},
\label{eq:EnvPhiInequality}
\end{align}
indicating that $E_{n,v}^{\text{pw}}[\{\chi_{n,p}, \lambda_{n,p}\}]$ reaches the minimum $E_{n,v}$ when $\rho_{\{\chi_{n,p}, \lambda_{n,p}\}} = \rho_v$ and equals $E_v$. \(\square\)

Theorem 3 thus demonstrates that $\{\chi_{n,p}, \lambda_{n,p}\}$ fundamentally defines the system, with the ground state energy obtainable by minimizing the functional $E_{n,v}^{\text{pw}}[\{\chi_{n,p}, \lambda_{n,p}\}]$. For simplicity, $E_{n,v}^{\text{pw}}[\{\chi_{n,p}, \lambda_{n,p}\}]$ is decomposed as:
\begin{align}
E&_{n,v}^{\text{pw}}[\{\chi_{n,p}, \lambda_{n,p}\}] = T_{n}[\{\chi_{n,p}, \lambda_{n,p}\}] + J[\rho_{\{\chi_{n,p}, \lambda_{n,p}\}}] \nonumber \\
&+ E_{n,xc}[\{\chi_{n,p}, \lambda_{n,p}\}] + \int v(\textbf{r}) \rho_{\{\chi_{n,p}, \lambda_{n,p}\}}(\textbf{r}) \, d\textbf{r},
\label{eq:Envdecomp}
\end{align}
where $T_n$ represents the kinetic energy \cite{Su2021pra},
\begin{equation}
T_{n}[\{\chi_{n,p}, \lambda_{n,p}\}] = - \sum_{p} \lambda_{n,p} \langle \chi_{n,p} | \nabla^2 | \chi_{n,p} \rangle.
\label{eq:Tnchi}
\end{equation}
$J$ denotes the Coulomb energy, $E_{n,xc}$ is the exchange-correlation energy accounting for the remaining kinetic and electron interaction energies not covered by $T_n$ and $J$, and the final term represents the external potential energy.

The above derivations and conclusions require only minor modifications to apply to $\Phi_n$ and $\{\varphi_{n,i}\}$. Specifically, in Eqs. \ref{eq:Env}-\ref{eq:Tnchi}, replace $\{\chi_{n,p}, \lambda_{n,p}\}$ with $\Phi_n$ and $\{\varphi_{n,i}\}$ respectively, substituting $\rho_{\{\chi_{n,p}, \lambda_{n,p}\}}$ with $\rho_{\Phi_n} = \langle \Phi_n | \hat{\rho} | \Phi_n \rangle$ and $\rho_{\{\varphi_{n,i}\}} =2 \sum_i |\varphi_{n,i}(\mathbf{r})|^2$. The energy functionals  \(E_{n,v}^{\text{pw}}[\Phi_n]\) and \(E_{n,v}^{\text{pw}}[\{\varphi_{n,i}\}]\) follow the decomposition in Eq. \ref{eq:Envdecomp}, with \(T_n\) of Eq. \ref{eq:Tnchi}, expressed respectively in terms of $\Phi_n$ and \(\{\varphi_{n,i}\}\), given by\cite{Su2021pra}:
\begin{equation}
T_{n}[\Phi_n] = \langle \Phi_n | \hat{T} | \Phi_n \rangle,
\label{eq:Tnphi}
\end{equation}
\begin{equation}
T_{n}[\{\varphi_{n,i}\}] = - \sum_{i} \langle \varphi_{n,i} | \nabla^2 | \varphi_{n,i} \rangle.
\label{eq:Tnvarphi}
\end{equation}

Further clarifications on the theorems are provided here. While current discussions primarily address pure ground states of interacting systems, extensions to ensemble states are feasible. This involves generalizing \(\rho_v\) to ensemble IVR densities and redefining energy functionals in Eqs. \ref{eq:Levy} and \ref{eq:EnvPhi}, as detailed in Ref. \citenum{Valone1980jcp}. Specifically, for a \(g\)-fold ensemble state, Eq. \ref{eq:EnvPhi} is reformulated as:
\begin{equation}
E_{n,v}^{\text{pw}}[\{\chi_{n,p}, \lambda_{n,p}\}] = \min_{\{\Psi^k,d^k\} \rightarrow \rho_{\{\chi_{n,p}, \lambda_{n,p}\}}} \sum_{k=1}^g d^k \langle \Psi^k | \hat{H}_v | \Psi^k \rangle,
\label{eq:eEnvchi}
\end{equation}
where the search for \(g\) anti-symmetric wavefunctions \(\{\Psi^k\}\) (each with its density \(\rho^k\)) and their weights \(\{d^k\}\) (satisfying \(0 \leq d^k \leq 1\) and \(\sum_{k=1}^g d^k = 1\)) ensures the ensemble density matches \(\rho_{\{\chi_{n,p}, \lambda_{n,p}\}}\), i.e., $\sum_{k=1}^g d^k \rho^k(\mathbf{r}) = \rho_{\{\chi_{n,p}, \lambda_{n,p}\}}$. Concerning \(g\), by assigning it a sufficiently large number, non-contributing weights will reduce to zero during the minimization process, thereby effectively capturing the system's true degeneracy. The energy functionals \(E_{n,v}^{\text{pw}}[\Phi_n]\) and  \(E_{n,v}^{\text{pw}}[\{\varphi_{n,i}\}]\) are adapted similarly. Notably, even degenerate systems managed with a single Slater determinant \(\Phi_n\) benefit from using hypercomplex orbitals \(\{\varphi_{n,i}\}\), equating to the application of HCOs and their occupations \(\{\chi_{n,p}, \lambda_{n,p}\}\), thus effectively representing ensemble IVR densities and strong correlation effects \cite{Su2021pra,Su2022es,Zhang2023pra}.

Moreover, the theoretical framework can be extended to handle excited states, including ensemble excited states. According to the Levy constrained search method and its extension \cite{Levy1979pnas,Perdew1984prb,Gorling1999pra,yang2024foundation}, when the trial density corresponds to an excited state, it identifies a stationary point rather than the global minimum. Thus, by modifying the minimization process in Eqs. \ref{eq:Ev} and \ref{eq:Env} to target these stationary points, the theorems can be applied to compute the energies and densities of excited states.

While \(\Phi_n\), \(\{\varphi_{n,i}\}\), and \(\{\chi_{n,p}, \lambda_{n,p}\}\) all serve as fundamental variables of the system, only HCOs and their occupations, \(\{\chi_{n,p}, \lambda_{n,p}\}\), are real values \cite{Su2021pra}. Additionally, Theorem 1 indicates that \(\{\chi_{n,p}, \lambda_{n,p}\}\) possess the smallest domain, enhancing their operability and optimizations. Furthermore, the fractional occupations of HCOs can effectively capture strong correlation effects, offering advantages in functional development.
Notably, at \(n=0\), HCO occupations are strictly binary---either 1 or 0---clearly delineating occupied from unoccupied orbitals. This scenario allows the above three theorems to lay a rigorous theoretical foundation for various (occupied and unoccupied) orbital functionals \cite{Zhao2004jpca,Grimme2006jcp,Zhang2009pnas,Scuseria2008,Furche2008jcp,Hebelmann2011,Ren2011prl,van2013pra,Cornaton2014ijqc,Su2016,Tahir2019prb,Zhang2019jpcl} and corresponding orbital optimization \cite{Yaffe1976pra,Lochan2007jcp,Pederson2014jcp,Neese2009jctc,Peverati2013jcp,Jin2017jpcl} implementations. In fact, even at \(n=0\), this approach can circumvent many issues associated with KSVR densities, thereby explaining the favorable outcomes observed in existing orbital-optimization calculations \cite{Lochan2007jcp,Pederson2014jcp,Neese2009jctc,Peverati2013jcp,Jin2017jpcl,Chen2017jpcl}.
When \(n=\infty\), the constraints on \(\{\chi_{n,p}, \lambda_{n,p}\}\) imposed by Eqs. \ref{eq:cons_hco} and \ref{eq:cons_hcoo} align precisely with the ensemble \(N\)-representability constraints on natural orbitals and their occupations, thereby enabling these theorems to demonstrate the rigor of reduced density matrix functional theory \cite{Gilbert1975prb,Levy1979pnas,Valone1980jcp,Cioslowski2020jcp,Liebert2023jcp} from another perspective.

The above addresses the spin-compensated case of \(N/2\) electrons for each spin. As for spin-polarized systems containing \(N_\alpha\) and \(N_\beta\) electrons with differing spins, \(\Phi_n\) then comprises \(N_\alpha\) orbitals \(\{\varphi_{n,i}^\alpha\}\) and \(N_\beta\) orbitals \(\{\varphi_{n,i}^\beta\}\), each linked to distinct sets of HCOs and their occupations, specifically \(\{\chi_{n,p}^\alpha, \lambda_{n,p}^\alpha\}\) and \(\{\chi_{n,p}^\beta, \lambda_{n,p}^\beta\}\). In this scenario, the densities---including \(\rho_v\), \(\rho_{w_s}\), and \(\rho_n\)---are accordingly decomposed into components for \(\alpha\) and \(\beta\) spins. Consequently, similar derivations can be carried out for both spins, details of which are not repeated here.

In summary, this study integrates various functionals---those dependent on occupied, unoccupied, or fractionally-occupied orbitals---into a unified framework. Each type of functional corresponds to a Clifford algebra with a specific dimensional basis, with their theoretical rigor thoroughly demonstrated. Specifically, Theorem 3 establishes a variational principle that lays a robust theoretical foundation for both orbital optimization and simultaneous orbital and occupation optimization. Furthermore, Theorem 2 investigates the domains of densities determined by the basis dimensions within Clifford algebras, confirming that these methodologies can effectively overcome the limitations posed by the KSVR assumption. Moreover, when properly implemented, these methodologies can also resolve common issues encountered in GGA and hybrid functional calculations, particularly when the desired solutions incorrectly target excited states of noninteracting systems---this work ensures the validity of such implementation.

%
%

Support from the National Natural Science Foundation of China (Grants No. 22122303 and No. 22073049) and Fundamental Research Funds for the Central Universities (Nankai University, Grant No. 63206008) is appreciated.

\appendix
\section{Details of the proof of Theorem 1}

\textbf{Theorem 1:} The mappings among $\Phi_n$, $\{\varphi_{n,i}\}$, and $\{\chi_{n,p}, \lambda_{n,p}\}$ are surjective for a given $n$, as shown in:
\begin{equation}
\label{seq:domain}
\Phi_n \Leftrightarrow\{\varphi_{n,i}\} \Leftrightarrow \{\chi_{n,p}, \lambda_{n,p}\}.
\end{equation}
Specifically, there exists a one-to-one mapping between $\Phi_n$ and $\{\varphi_{n,i}\}$, and a many-to-one mapping between $\{\varphi_{n,i}\}$ and $\{\chi_{n,p}, \lambda_{n,p}\}$, within the unitary rotation of $\{\varphi_{n,i}\}$ and of $\{\chi_{n,p}\}$ with the same occupation. 

\textbf{Proof 1:} The one-to-one surjective mapping between $\Phi_n$ and $\{\varphi_{n,i}\}$, within the unitary rotation of $\{\varphi_{n,i}\}$, is evident. The demonstration below illustrates the many-to-one surjective mapping between $\{\varphi_{n,i}\}$ and $\{\chi_{n,p}, \lambda_{n,p}\}$, within the unitary rotation of $\{\varphi_{n,i}\}$ and of $\{\chi_{n,p}\}$ with the same occupation.

For completeness, a brief review of relevant concepts is provided. Hypercomplex orbitals are defined as \cite{Su2021pra}
\begin{equation}
\label{seq:varphi}
\varphi_{n,p}(\mathbf{r}) = \phi_{p}^{0}(\mathbf{r}) + \sum_{\mu=1}^{n} \phi_{p}^{\mu}(\mathbf{r}) e_\mu,
\end{equation}
where $\{e_1, e_2, \ldots, e_n\}$ form a basis in a Clifford algebra of dimension $n$ \cite{CA2019}. For constructing $\Phi_n$, $N/2$ hypercomplex orbitals are utilized; without loss of generality, the first $N/2$ orbitals, denoted $\{\varphi_{n,i}\}$, are selected. The set $\{\phi_{p}^{\mu}\}$ consists of real functions expanded on a set of orthonormal functions $\{\xi_p\}$, which does not compromise generality:
\begin{equation}
\label{seq:phi}
\phi_p^\mu(\textbf{r}) = \sum_{q=1}^K \xi_q(\textbf{r}) V_{pq}^\mu,
\end{equation}
where $K$ is the dimension of the basis set, and $\{{V}^{\mu}\}$ are a set of $n+1$ $K \times K$ matrices, each ${V}^{\mu}$ representing the expansion coefficients for the $\mu$-th component of the hypercomplex orbitals. These matrices preserve the orthonormality of the hypercomplex orbitals, leading to the following conditions on $\{{V}^{\mu}\}$:
\begin{align}
\label{seq:cond_hc}
\begin{cases}
\sum_{\mu=0}^{n} {V}^{\mu}{V}^{\mu T} = {I}_K,  \\
\sum_{\mu=0}^{n} {V}^{\mu T}{V}^{\mu} = {I}_K,  \\
{V}^{\mu}{V}^{\nu T} = {V}^{\nu}{V}^{\mu T},  \\
{V}^{\mu T}{V}^{\nu} = {V}^{\nu T}{V}^{\mu},
\end{cases}
\end{align}
where ${I}_K$ is the $K \times K$ identity matrix, and the superscript $T$ denotes the matrix transpose.
Through a series of derivations \cite{Su2021pra}, the density can be represented on the set $\{\chi_{n,p}, \lambda_{n,p}\}$, as given by:
\begin{equation}
\rho_n(\mathbf{r}) = 2 \sum_p \lambda_{n,p} |\chi_{n,p}(\mathbf{r})|^2.
\label{seq:trialrho}
\end{equation}
Here, $\{\chi_{n,p}, \lambda_{n,p}\}$ correspond to the eigenvectors and eigenvalues, respectively, of the symmetric matrix $D_n$, defined by:
\begin{equation}
\label{seq:D_hc}
D_n = \sum_{\mu=0}^{n} {V}^{\mu T} {\rm{I}}^{N/2}_K {V}^{\mu},
\end{equation}
where ${\rm{I}}^{N/2}_K$ is a $K \times K$ diagonal matrix, with the first $N/2$ diagonal entries set to 1, and the rest set to 0. The spectral decomposition of $D_n$ is given by:
\begin{equation}
\label{seq:D_de}
D_n = U \Lambda U^T.
\end{equation}
where $\Lambda$ is a diagonal matrix containing the eigenvalues $\{\lambda_{n,p}\}$, and $U$, a unitary matrix, defines $\{\chi_{n,p}\}$ as $\chi_{n,p} = \sum_{q=1}^K \xi_q U_{qp}$. Thus, $\{\chi_{n,p}\}$ are orthonormal:
\begin{equation}
\label{seq:cons_hco}
\langle \chi_{n,p} | \chi_{n,q} \rangle = \delta_{pq}.
\end{equation}
Interestingly, the constraints on $\{\lambda_{n,p}\}$ depend on the dimension $n$ of the Clifford algebra. Specifically, for any $n$, all $\{{V}^{\mu}\}$ that satisfy Eq. \ref{seq:cond_hc} define the domain $\Omega_n$ for $\{\chi_{n,p}, \lambda_{n,p}\}$. Notably, at $n=0$, $\Omega_{n=0}$ encompasses $\{\chi_{n,p}\}$ that meet Eq. \ref{seq:cons_hco}, with $N/2$ of them having an occupation of 1 and the rest 0, thereby clearly delineating occupied from unoccupied orbitals. For further discussion, see Refs. \citenum{Su2021pra} and \citenum{Zhang2023pra}.

The surjectivity of the mapping between $\{\varphi_{n,i}\}$ and $\{\chi_{n,p}, \lambda_{n,p}\}$ is first established. For any orthonormal set $\{\varphi_{n,i}\}$, a corresponding set $\{{V}^{\mu}\}$ exists that satisfies Eq. \ref{seq:cond_hc}, thus ensuring the presence of $\{\chi_{n,p}, \lambda_{n,p}\}$ in $\Omega_n$. Conversely, for each $\{\chi_{n,p}, \lambda_{n,p}\}$ in $\Omega_n$, there must be a set $\{{V}^{\mu}\}$ compliant with Eq. \ref{seq:cond_hc}, defining a compatible set $\{\varphi_{n,i}\}$. Without such a set $\{{V}^{\mu}\}$, $\{\chi_{n,p}, \lambda_{n,p}\}$ would not fall within $\Omega_n$. This confirms the surjective nature of the mapping between $\{\varphi_{n,i}\}$ and $\{\chi_{n,p}, \lambda_{n,p}\}$.

For any unitary rotation of the first $N/2$ hypercomplex orbitals $\{\varphi_{n,i}\}$, it corresponds to the following transformation on all matrices in $\{{V}^{\mu}\}$:
\begin{equation}
\label{seq:Vmup}
V'^\mu = W V^\mu,
\end{equation}
where $W$ is defined as:
\begin{equation}
\label{seq:O_hc}
{\rm{W}}=\begin{bmatrix}
Q & O_{N/2, K-N/2} \\
O_{K-N/2, N/2} & I_{K-N/2}
\end{bmatrix},
\end{equation}
with $Q$ being a $N/2 \times N/2$ unitary matrix, and $O_{I,J}$ denoting an $I \times J$ zero matrix. It can be derived that for any $\{{V}^{\mu}\}$ satisfying Eq. \ref{seq:cond_hc}, and its corresponding $D_n$ defined by Eq. \ref{seq:D_hc}, applying the transformation in Eq. \ref{seq:Vmup} to all matrices in $\{{V}^{\mu}\}$ yields a new set $\{{V'}^{\mu}\}$ that also satisfies Eq. \ref{seq:cond_hc}, producing the same $D_n$ as defined by Eq. \ref{seq:D_hc}. Therefore, the transformed and original matrices correspond to the same $\{\chi_{n,p}, \lambda_{n,p}\}$. Moreover, if $D_n$ has identical eigenvalues, i.e., multiple orbitals in $\{\chi_{n,p}\}$ have the same occupations, any unitary rotation among these orbitals, by the spectral decomposition defined in Eq. \ref{seq:D_de}, leads to the same $D_n$ before and after the transformation, thus corresponding to the same $\{{V}^{\mu}\}$ and their defined $\{\varphi_{n,i}\}$. In fact, the invariance under these transformations also applies to the energy functionals defined in this work. The mappings between \(\{\varphi_{n,i}\}\) and \(\{\chi_{n,p}, \lambda_{n,p}\}\) discussed below do not account for the changes these transformations impart to \(\{\varphi_{n,i}\}\) and \(\{\chi_{n,p}, \lambda_{n,p}\}\).

Lastly, the many-to-one mapping between \(\{\varphi_{n,i}\}\) and \(\{\chi_{n,p}, \lambda_{n,p}\}\) is confirmed. For a given set of \(\{\varphi_{n,i}\}\), there is a specific set of \(\{{V}^{\mu}\}\) that satisfies Eq. \ref{seq:cond_hc} and a distinct matrix \(D_n\), thus associating a unique set of \(\{\chi_{n,p}, \lambda_{n,p}\}\) with it. Additionally, if the roles of two bases in all hypercomplex orbitals (Eq. \ref{seq:varphi}), such as \(e_\mu\) and \(e_\nu\), and the functions \(\phi_{p}^{\mu}\) and \(\phi_{p}^{\nu}\) are interchanged, this is equivalent to swapping the coefficient matrices \({V}^{\mu}\) and \({V}^{\nu}\). Clearly, such a transformation cannot be achieved by a unitary rotation of \(\{\varphi_{n,i}\}\); however, it is evident that the new matrices \(\{{V'}^{\mu}\}\) also satisfy Eq. \ref{seq:cond_hc}, and the \(D_n\) defined by Eq. \ref{seq:D_hc} remains unchanged after the transformation. Therefore, the mapping between \(\{\varphi_{n,i}\}\) and \(\{\chi_{n,p}, \lambda_{n,p}\}\) is many-to-one. This discussion only mentions a simple transformation; there are evidently other more complex transformations involving hypercomplex orbitals that leave \(\{\chi_{n,p}, \lambda_{n,p}\}\) unchanged, hence the domain of \(\{\chi_{n,p}, \lambda_{n,p}\}\) is considerably smaller.  \(\square\)

\bibliographystyle{aip}
\bibliography{ref}

\begin{thebibliography}{10}

\bibitem{HK1964}
P.~Hohenberg and W.~Kohn,
\newblock Phys. Rev. {\bf 136}, B864 (1964).

\bibitem{Parr1978jcp}
R.~G. Parr, R.~A. Donnelly, M.~Levy, and W.~E. Palke,
\newblock J. Chem. Phys. {\bf 68}, 3801 (1978).

\bibitem{Katriel1981ijqc}
J.~Katriel, C.~J. Appellof, and E.~R. Davidson,
\newblock Int. J. Quant. Chem. {\bf 19}, 293 (1981).

\bibitem{Levy1982pra}
M.~Levy,
\newblock Phys. Rev. A {\bf 26}, 1200 (1982).

\bibitem{Englisch1983}
H.~Englisch and R.~Englisch,
\newblock Physica A {\bf 121}, 253 (1983).

\bibitem{KS1965}
W.~Kohn and L.~J. Sham,
\newblock Phys. Rev. {\bf 140}, A1133 (1965).

\bibitem{Levy1979pnas}
M.~Levy,
\newblock Proc. Natl. Acad. Sci. USA {\bf 76}, 6062 (1979).

\bibitem{Valone1980jcp}
S.~M. Valone,
\newblock J. Chem. Phys. {\bf 73}, 4653 (1980).

\bibitem{Lieb1983ijqc}
E.~H. Lieb,
\newblock Int. J. Quant. Chem. {\bf 24}, 243 (1983).

\bibitem{Yang2004prl}
W.~Yang, P.~W. Ayers, and Q.~Wu,
\newblock Phys. Rev. Lett. {\bf 92}, 146404 (2004).

\bibitem{Talman1976pra}
J.~D. Talman and W.~F. Shadwick,
\newblock Phys. Rev. A {\bf 14}, 36 (1976).

\bibitem{KLI1992pra}
J.~B. Krieger, Y.~Li, and G.~J. Iafrate,
\newblock Phys. Rev. A {\bf 45}, 101 (1992).

\bibitem{Zhao1994pra}
Q.~Zhao, R.~C. Morrison, and R.~G. Parr,
\newblock Phys. Rev. A {\bf 50}, 2138 (1994).

\bibitem{van1994pra}
R.~van Leeuwen and E.~J. Baerends,
\newblock Phys. Rev. A {\bf 49}, 2421 (1994).

\bibitem{Ivanov1999prl}
S.~Ivanov, S.~Hirata, and R.~J. Bartlett,
\newblock Phys. Rev. Lett. {\bf 83}, 5455 (1999).

\bibitem{Gorling1999prl}
A.~G\"orling,
\newblock Phys. Rev. Lett. {\bf 83}, 5459 (1999).

\bibitem{Yang2002prl}
W.~Yang and Q.~Wu,
\newblock Phys. Rev. Lett. {\bf 89}, 143002 (2002).

\bibitem{Wu2003jcp}
Q.~Wu and W.~Yang,
\newblock J. Chem. Phys. {\bf 118}, 2498 (2003).

\bibitem{Kummel2003prl}
S.~K\"ummel and J.~P. Perdew,
\newblock Phys. Rev. Lett. {\bf 90}, 043004 (2003).

\bibitem{Bulat2007jcp}
F.~A. Bulat, T.~Heaton-Burgess, A.~J. Cohen, and W.~Yang,
\newblock J. Chem. Phys. {\bf 127}, 174101 (2007).

\bibitem{Lee2019prl}
J.~Lee, L.~W. Bertels, D.~W. Small, and M.~Head-Gordon,
\newblock Phys. Rev. Lett. {\bf 123}, 113001.

\bibitem{Jin2020fd}
Y.~Jin, N.~Q. Su, Z.~Chen, and W.~Yang,
\newblock Faraday Discuss. {\bf 224}, 9 (2020).

\bibitem{Yu2021jcp}
J.~M. Yu, B.~D. Nguyen, J.~Tsai, D.~J. Hernandez, and F.~Furche,
\newblock J. Chem. Phys. {\bf 155}, 040902 (2021).

\bibitem{Su2021pra}
N.~Q. Su,
\newblock Phys. Rev. A {\bf 104}, 052809 (2021).

\bibitem{DFT2022PCCP}
A.~M. Teale et~al.,
\newblock Phys. Chem. Chem. Phys. {\bf 24}, 28700 (2022).

\bibitem{B88}
A.~D. Becke,
\newblock Phys. Rev. A {\bf 38}, 3098 (1988).

\bibitem{LYP}
C.~Lee, W.~Yang, and R.~G. Parr,
\newblock Phys. Rev. B {\bf 37}, 785 (1988).

\bibitem{PBE}
J.~P. Perdew, K.~Burke, and M.~Ernzerhof,
\newblock Phys. Rev. Lett. {\bf 77}, 3865 (1996).

\bibitem{Sun2015prl}
J.~Sun, A.~Ruzsinszky, and J.~P. Perdew,
\newblock Phys. Rev. Lett. {\bf 115}, 036402 (2015).

\bibitem{LYPr}
W.~Ai, W.-H. Fang, and N.~Q. Su,
\newblock J. Phys. Chem. Lett. {\bf 12}, 1207 (2021).

\bibitem{Becke1993jcp}
A.~D. Becke,
\newblock J. Chem. Phys. {\bf 98}, 5648 (1993).

\bibitem{Adamo1999jcp}
C.~Adamo and V.~Barone,
\newblock J. Chem. Phys. {\bf 110}, 6158 (1999).

\bibitem{Ernzerhof1999jcp}
M.~Ernzerhof and G.~E. Scuseria,
\newblock J. Chem. Phys. {\bf 110}, 5029 (1999).

\bibitem{Xu2004}
X.~Xu and W.~A. Goddard,
\newblock J. Chem. Phys. {\bf 121}, 4068 (2004).

\bibitem{Zhao2008acr}
Y.~Zhao and D.~G. Truhlar,
\newblock Acc. Chem. Res. {\bf 41}, 157 (2008).

\bibitem{Iikura2001}
H.~Iikura, T.~Tsuneda, T.~Yanai, and K.~Hirao,
\newblock J. Chem. Phys. {\bf 115}, 3540 (2001).

\bibitem{Savin1996}
A.~Savin,
\newblock On degeneracy, near-degeneracy and density functional theory,
\newblock in {\em Recent Developments and Applications of Modern Density
  Functional Theory}, edited by J.~Seminario, volume~4 of {\em Theoretical and
  Computational Chemistry}, pages 327--357, Elsevier, 1996.

\bibitem{Yanai2004}
T.~Yanai, D.~P. Tew, and N.~C. Handy,
\newblock Chem. Phys. Lett. {\bf 393}, 51 (2004).

\bibitem{Chai2008jcp}
J.-D. Chai and M.~Head-Gordon,
\newblock J. Chem. Phys. {\bf 128}, 084106 (2008).

\bibitem{HSE}
J.~Heyd, G.~E. Scuseria, and M.~Ernzerhof,
\newblock J. Chem. Phys. {\bf 118}, 8207 (2003).

\bibitem{Garrick2020prx}
R.~Garrick, A.~Natan, T.~Gould, and L.~Kronik,
\newblock Phys. Rev. X {\bf 10}, 021040 (2020).

\bibitem{Zhao2004jpca}
Y.~Zhao, B.~J. Lynch, and D.~G. Truhlar,
\newblock J. Phys. Chem. A {\bf 108}, 4786 (2004).

\bibitem{Grimme2006jcp}
S.~Grimme,
\newblock J. Chem. Phys. {\bf 124}, 034108 (2006).

\bibitem{Zhang2009pnas}
Y.~Zhang, X.~Xu, and W.~A. Goddard,
\newblock Proc. Natl. Acad. Sci. USA {\bf 106}, 4963 (2009).

\bibitem{Scuseria2008}
G.~E. Scuseria, T.~M. Henderson, and D.~C. Sorensen,
\newblock J. Chem. Phys. {\bf 129}, 231101 (2008).

\bibitem{Furche2008jcp}
F.~Furche,
\newblock J. Chem. Phys. {\bf 129}, 114105 (2008).

\bibitem{Hebelmann2011}
A.~He{\ss}elmann and A.~G\"orling,
\newblock Mol. Phys. {\bf 109}, 2473 (2011).

\bibitem{Ren2011prl}
X.~Ren, A.~Tkatchenko, P.~Rinke, and M.~Scheffler,
\newblock Phys. Rev. Lett. {\bf 106}, 153003 (2011).

\bibitem{van2013pra}
H.~van Aggelen, Y.~Yang, and W.~Yang,
\newblock Phys. Rev. A {\bf 88}, 030501(R) (2013).

\bibitem{Cornaton2014ijqc}
Y.~Cornaton and E.~Fromager,
\newblock Int. J. Quant. Chem. {\bf 114}, 1199 (2014).

\bibitem{Su2016}
N.~Q. Su and X.~Xu,
\newblock WIREs Comput. Mol. Sci. {\bf 6}, 721 (2016).

\bibitem{Tahir2019prb}
M.~N. Tahir and X.~Ren,
\newblock Phys. Rev. B {\bf 99}, 195149 (2019).

\bibitem{Zhang2019jpcl}
I.~Y. Zhang and X.~Xu,
\newblock J. Phys. Chem. Lett. {\bf 10}, 2617 (2019).

\bibitem{Muller1984rpa}
A.~Muller,
\newblock Phys. Lett. A {\bf 105}, 446 (1984).

\bibitem{GU1998prl}
S.~Goedecker and C.~J. Umrigar,
\newblock Phys. Rev. Lett. {\bf 81}, 866 (1998).

\bibitem{Gritsenko2005jcp}
O.~Gritsenko, K.~Pernal, and E.~J. Baerends,
\newblock J. Chem. Phys. {\bf 122}, 204102 (2005).

\bibitem{Sharma2008prb}
S.~Sharma, J.~K. Dewhurst, N.~N. Lathiotakis, and E.~K.~U. Gross,
\newblock Phys. Rev. B {\bf 78}, 201103(R) (2008).

\bibitem{Piris2021prl}
M.~Piris,
\newblock Phys. Rev. Lett. {\bf 127}, 233001 (2021).

\bibitem{Ai2022jpcl}
W.~Ai, W.-H. Fang, and N.~Q. Su,
\newblock J. Phys. Chem. Lett. {\bf 13}, 1744 (2022).

\bibitem{Gibney2023prl}
D.~Gibney, J.-N. Boyn, and D.~A. Mazziotti,
\newblock Phys. Rev. Lett. {\bf 131}, 243003 (2023).

\bibitem{Chai2012jcp}
J.-D. Chai,
\newblock J. Chem. Phys. {\bf 136}, 154104 (2012).

\bibitem{Yaffe1976pra}
L.~G. Yaffe and W.~A. Goddard,
\newblock Phys. Rev. A {\bf 13}, 1682 (1976).

\bibitem{Lochan2007jcp}
R.~C. Lochan and M.~Head-Gordon,
\newblock J. Chem. Phys. {\bf 126}, 164101 (2007).

\bibitem{Pederson2014jcp}
M.~R. Pederson, A.~Ruzsinszky, and J.~P. Perdew,
\newblock J. Chem. Phys. {\bf 140}, 121103 (2014).

\bibitem{Neese2009jctc}
F.~Neese, T.~Schwabe, S.~Kossmann, B.~Schirmer, and S.~Grimme,
\newblock J. Chem. Theory Comput. {\bf 5}, 3060 (2009).

\bibitem{Peverati2013jcp}
R.~Peverati and M.~Head-Gordon,
\newblock J. Chem. Phys. {\bf 139}, 024110 (2013).

\bibitem{Jin2017jpcl}
Y.~Jin, D.~Zhang, Z.~Chen, N.~Q. Su, and W.~Yang,
\newblock J. Phys. Chem. Lett. {\bf 8}, 4746 (2017).

\bibitem{Piris2009jcc}
M.~Piris and J.~M. Ugalde,
\newblock J. Comput. Chem. {\bf 30}, 2078 (2009).

\bibitem{Yao2021jpcl}
Y.-F. Yao, W.-H. Fang, and N.~Q. Su,
\newblock J. Phys. Chem. Lett. {\bf 12}, 6788 (2021).

\bibitem{Yao2024arXiv}
Y.-F. Yao and N.~Q. Su,
\newblock Enhancing reduced density matrix functional theory calculations by
  coupling orbital and occupation optimizations, 2024,
\newblock arXiv:2402.03532.

\bibitem{Cartier2024arXiv}
N.~G. Cartier and K.~J.~H. Giesbertz,
\newblock Exploiting the hessian for a better convergence of the scf rdmft
  procedure, 2024,
\newblock arXiv:2401.16324.

\bibitem{PY1989}
R.~G. Parr and W.~Yang,
\newblock {\em Density-Functional Theory of Atoms and Molecules},
\newblock Oxford University Press: New York, 1989.

\bibitem{CA2019}
J.~{Vaz Jr.} and R.~da~{Rocha Jr.},
\newblock {\em An Introduction to Clifford Algebras and Spinors},
\newblock Oxford University Press: New York, 2019.

\bibitem{Su2022es}
N.~Q. Su,
\newblock Electron. Struct. {\bf 4}, 014011 (2022).

\bibitem{footnote}
See the Supplemental Material for details of the proof of Theorem 1.

\bibitem{Zhang2023pra}
T.~Zhang and N.~Q. Su,
\newblock Phys. Rev. A {\bf 108}, 052801 (2023).

\bibitem{Harriman1981pra}
J.~E. Harriman,
\newblock Phys. Rev. A {\bf 24}, 680 (1981).

\bibitem{Perdew1984prb}
J.~P. Perdew and M.~Levy,
\newblock Phys. Rev. B {\bf 31}, 6264 (1985).

\bibitem{Gorling1999pra}
A.~G\"orling,
\newblock Phys. Rev. A {\bf 59}, 3359 (1999).

\bibitem{yang2024foundation}
W.~Yang and P.~W. Ayers,
\newblock Foundation for the $\delta$scf approach in density functional theory,
  2024,
\newblock arXiv:2403.04604.

\bibitem{Chen2017jpcl}
Z.~Chen, D.~Zhang, Y.~Jin, Y.~Yang, N.~Q. Su, and W.~Yang,
\newblock J. Phys. Chem. Lett. {\bf 8}, 4479 (2017).

\bibitem{Gilbert1975prb}
T.~L. Gilbert,
\newblock Phys. Rev. B {\bf 12}, 2111 (1975).

\bibitem{Cioslowski2020jcp}
J.~Cioslowski,
\newblock J. Chem. Phys. {\bf 153}, 154108 (2020).

\bibitem{Liebert2023jcp}
J.~Liebert, A.~Y. Chaou, and C.~Schilling,
\newblock J. Chem. Phys. {\bf 158}, 214108 (2023).

\end{thebibliography}
 
\end{document}